\documentclass[12pt,fleqn]{article}

\usepackage{graphicx}
\usepackage{dcolumn}
\usepackage{bm}
\usepackage{amsmath}
\usepackage{amssymb}  
\usepackage{subfigure}
\usepackage{psfig,float}
\def\OMIT#1{}
\begin{document} 

\title{A stochastic approach to multi-gene \\
expression dynamics}

\author{T. Ochiai{\footnote{These authors contributed equally to this work. \newline
Corresponding authors: ochiai@kuicr.kyoto-u.ac.jp, nacher@kuicr.kyoto-u.ac.jp
}}, J.C. Nacher{\footnotemark[1]}, T. Akutsu}     
 
\maketitle
 
\begin{center}
{\it Bioinformatics Center, Institute for Chemical Research, Kyoto University, }\end{center}
\begin{center}
{\it Uji, 611-0011, Japan}
\end{center}

\begin{center}
\end{center} 

\begin{center}

\end{center}

\begin{abstract}
{\small{

In the last years, tens of thousands gene expression profiles for cells of several organisms have been monitored. Gene expression 
is a complex transcriptional process where {\it mRNA} molecules are translated into proteins, which control most of the cell functions.
In this process, the correlation among genes is crucial to determine the specific functions of genes. Here, we propose a
novel multi-dimensional stochastic approach to deal with the gene correlation phenomena. Interestingly, our stochastic 
framework suggests that the study of the gene correlation requires only one theoretical assumption -{\it Markov property-} 
and the experimental transition probability, which characterizes the gene correlation system. Finally, a gene expression 
experiment is proposed for future applications of the model.
}}
\end{abstract}


\section{Introduction}
In a living organism or cell the collective behavior of thousands of 
genes and their products (for example mRNA and proteins) are embedded 
in a complex architecture that creates the mystery of life. More than 
10 years ago, methods in molecular biology worked on a "{\it one gene- one experiment}" framework, 
meaning that the throughput is constrained to only one gene and therefore, the whole image of gene function 
is difficult to visualize. An emerging technology called {\it DNA} microarray/GeneChips \cite{hart,david} 
appeared in the recent years, attracting the interests among biologists, computer scientists, mathematicians and 
physicists. This 
technology allows us to monitor the whole transcribed genome on a single chip and offers the possibility to 
capture the correlations among thousands of expressed genes simultaneously.

In order to understand how the organism works, it is necessary 
to know {\it which} genes are expressed, {\it when} they are expressed 
and {\it how fast} they do. Gene expression is regulated by means of the gene regulation architecture 
system of the cells, which involves network of interactions among {\it DNA}, {\it mRNA}, proteins 
and hundreds of small ubiquitous molecules. These interactions involve 
many elements and different and complex mechanisms, therefore an intuitive 
understanding of the underlying dynamics is not easy to obtain. This is also true for the issue
 of gene correlation, which is the main aim of this letter. In particular, although many 
approaches and techniques, as for example Boolean networks, graph theory and 
control theory, have been used successfully in many cases, they are still far to achieve a general 
description of the dynamics of the regulation and correlation among genes. 


To shed light on this issue, here we 
propose a new theoretical model to deal 
with multi-gene correlation dynamics based on only one 
assumption: {\it Markov property} . Our approach will use the most 
general multi-variate stochastic process in order to obtain 
predictions about the correlations among genes.

In a previous work \cite{ochi}, we proposed a constructive approach to gene expression dynamics, which re-builds the
scale-free organization of genes (i.e., expression level $k$ decays as a power-law $k^{-\gamma}$ \cite{li,brass}) 
observed in recent experiments \cite{kuznetsov,kaneko,ueda}. There, we proposed a stochastic 
approach by assuming the Markov property
and by using the observed experimental transition probability data, which characterize the gene expression system. Although 
our companion paper \cite{ochi} 
succeeded to re-build the scale-free distribution, it may not provide much information about gene correlation 
phenomena because by construction it is one gene approach-like.
Therefore, here our aim is to exploit the novelty of our previous constructive approach by extending that one-dimensional model, 
to a multi-dimensional analysis (i.e., multi-gene correlation).

Our approach is based on two fundamental aspects: Markov property and stochastic process.

{\bf Markov property.} If we say that the system has a Markov property, we mean that the future is governed by the present and 
does not depend on the past. Our model assumes 
 Markov property to describe the multi-gene expression dynamics. Certainly, the living organisms are systems 
 with long term memory, and they are complex
 systems with large number of elements and interactions. However, from a physical point of view, although 
 we may not know all the variables in real situations, it may be enough to find a reduced number of variables whose behavior
  in time can be described as a Markovian process. Therefore, in our study we may assume that the most 
  relevant degree of freedom of the system is the gene expression level, and consequently, our gene 
  system has the Markov property.

{\bf Stochastic process.} Although many complex systems may be governed 
by non-stochastic processes, in the gene expression problem the random variation is reasonable,
plays a relevant role in cellular process, and furthermore stochastic noise have recently been measured and studied theoretically \cite{elo,
paulsson, blake, hasty}. For example, 
the expression level of thousands genes is very 
low, which
creates intrinsic uncertainties in the number of expressed genes in the cells \cite{kuznetsov}. Furthermore, we can even 
distinguish between inherent stochasticity ({\it intrinsic noise}) and external stochasticity ({\it extrinsic noise}). While 
the origin of the first one are the biochemical processes, and motivates that two identical 
genes become uncorrelated due to that randomness, the second one
represents sources of extrinsic noise, which change from cell to cell (i.e., fluctuations in elements among cells) \cite{elo, sato}.
Moreover, the number 
of molecules which are involved in signal transduction pathways fluctuates
from $10^2$ to $10^4$. Therefore, the randomness connected with elementary molecular interactions and their amplification 
in the signaling cascade generates significant spatio-temporal noise. Therefore, the stochastic approach is justified, and it seems more
appropriate and plausible than a deterministic approach. Finally, it is also worth reminding that the current experimental
 techniques also provide an additional source of fluctuation, which come from the ubiquitous instrumental noise
  (which may be around $30\%$ or more) from chip to chip with the current GeneChips technologies. 

On the other hand, one drawback of our approach is that 
the current existing experimental data of gene expression time 
series gene lacks of enough statistics. For example, in 
yeast \cite{yeast} and human \cite{human} organism experiments,
they analyzed the fluctuations in time
of many genes simultaneously (more than 30.000 in the case of human organism) by carrying out only {\it one} experiment

However, in order to have enough statistics, we believe that by 
using many experiments of gene chips (i.e., {\it many} experiments measure repeatedly fluctuations of
{\it many genes} in time under the same conditions), 
we may achieve a better 
understanding of the global nature and dynamics of gene correlation. Therefore, the theoretical approach proposed in this letter may be 
a useful guideline for such kind of experiments, and moreover may encourage them.

The paper is organized as follows. Section 2 describes the 
theoretical background, explains our proposed model for 
multi-gene correlations and presents the results of our simulated data. Section 3 explains the experimental 
proposal compatible with our theoretical study, and finally Section 4 presents the conclusions.

\section{ Methods and Result}
\subsection{Methods}

\subsubsection{Markov property and Differential Chapman-Kolmogorov Equation}

\paragraph{Markov property.}
We use multi-dimensional stochastic process for describing the multi-gene correlation dynamics \cite{Kampen,Wong,Gard,black}. 
Let $\{{\bf X}_t = X_t^1, \cdots,  X_t^N), 0 \le t <\infty \}$ be a multi-dimensional stochastic process.  
For $(t_n>\cdots>t_0)$,  the conditional probability density function
\begin{eqnarray*} 
p({\bf x}_n,t_n|{\bf x}_{n-1},t_{n-1};\cdots;{\bf x}_0,t_0)=p({\bf X}_{t_n}={\bf x}_n | {\bf X}_{t_{n-1}}={\bf x}_{n-1} ;\cdots; {\bf X}_{t_0}={\bf x}_0 )
\end{eqnarray*}
is defined as usual manner, where ${\bf x}=(x^1,\cdots,x^N)$ denotes the $N$ dimensional vector. It is said that 
a multi-dimensional stochastic process has "Markov property", when the condition
\begin{eqnarray}\label{eqn: Markov property}
p({\bf x}_n,t_n|{\bf x}_{n-1},t_{n-1};\cdots;{\bf x}_0,t_0)
=p({\bf x}_n,t_n|{\bf x}_{n-1},t_{n-1}) 
\end{eqnarray}
holds for arbitrary $t_n>\cdots>t_0$. In what follows, we assume that the probability 
density $p({\bf x},t|{\bf x}_0,t_0)$ has the time translation 
invariance $p({\bf x},t|{\bf x}_0,t_0)=p({\bf x},t+a|{\bf x}_0,t_0+a)$ for arbitrary $a$.

Our only one assumption is that {\it the multi-dimensional correlation dynamics of gene expression obeys the Markov property}. More 
precisely, we assume 
that the expression levels of each gene are denoted by the multi-dimensional stochastic process with Markov property ${\bf X}_t$.

\paragraph{Differential Chapman-Kolmogorov equation.}
For the matter of convenience, we write $p({\bf x},t)$ for $p({\bf x},t|{\bf x}_0,t_0)$. Then, if the multi-dimensional stochastic 
process has the Markov property (Eq. (\ref{eqn: Markov property})), the conditional probability density 
function $p({\bf x},t|{\bf x}_0,t_0)$ obeys the Differential Chapman-Kolmogorov equation valid for a system composed of $N$ genes and 
reads as follows:
\begin{eqnarray}\label{eqn: master equation}
\frac{\partial p({\bf x},t)}{\partial t}&=&-\sum_{i=1}^N\frac{\partial}{\partial x^i}\{a^i({\bf x})p({\bf x},t)\}+\frac{1}{2}\sum_{i,j=1}^N\frac{\partial^2}{\partial x^i x^j}\{b^{ij}({\bf x})p({\bf x},t)\}\nonumber\\
&&+\int d{\bf y}[W({\bf x}|{\bf y},t)p({\bf y},t)-W({\bf y}|{\bf x},t)p({\bf x},t)],
\end{eqnarray}
where the {\it drift} term $a^i({\bf x})$ is given by
\begin{eqnarray}\label{eqn: initial condition1}
\lim_{\epsilon \to 0}\frac{1}{\epsilon}\int_{|{\bf y}-{\bf x}|<\delta} (y^i-x^i)T_\epsilon({\bf y},{\bf x}) d{\bf y}=a^i({\bf x})+O(\delta),
\end{eqnarray}
and the {\it diffusion} matrix $b^{ij}({\bf x})$ reads as 
\begin{eqnarray}\label{eqn: initial condition2}
\lim_{\epsilon \to 0}\frac{1}{\epsilon}\int_{|{\bf y}-{\bf x}|<\delta} (y^i-x^i)(y^j-x^j) T_\epsilon({\bf y},{\bf x}) d{\bf y}=b^{ij}({\bf x})+O(\delta), 
\end{eqnarray}
and {\it jump} term $W({\bf y}|{\bf x},t)$ is given by
\footnote{Here we remark that although $W({\bf x}|{\bf y},t)=0$ seems 
to imply that $b^{ij}({\bf x})=0$, it is not correct since, in general, 
it is not possible to exchange the order of the limit and the integral in Eq. (\ref{eqn: initial condition2}). Therefore, it is possible that $b^{ij}({\bf x})$ is not zero, even if $W({\bf x}|{\bf y},t)=0$.}
\begin{eqnarray}\label{eqn: initial condition3}
W({\bf y}|{\bf x},t)=\lim_{\epsilon \to 0} T_\epsilon({\bf y},{\bf x}) / \epsilon.
\end{eqnarray}

Here $T_\epsilon({\bf y},{\bf x})$ is an Instantaneous Transition 
Probability (ITP) defined by $T_\epsilon({\bf y},{\bf x})=p({\bf y},t + \epsilon | {\bf x},t)$ for 
sufficiently small $\epsilon$. 

In the context of gene expression level, Eq. (\ref{eqn: master equation}) represents the 
dynamics of $N$ {\it mRNA} molecules (i.e., gene expression) in the cell. By using this equation, we can study 
the dynamics and correlation between genes $i$ and $j$. 

Next, we will explain three important processes of the Differential Chapman-Kolmogorov equation (\ref{eqn: master equation}) in the following paragraph.

\paragraph{Deterministic process.}
If the diffusion matrix $b^{ij}({\bf x})$ and the term $W({\bf x}|{\bf y},t)$ vanish, then the Differential Chapman-Kolomogorov 
equation (\ref{eqn: master equation}) is reduced to 
\begin{eqnarray}\label{eqn: determinisitic process}
\frac{\partial p({\bf x},t)}{\partial t}&=&-\sum_{i=1}^N\frac{\partial}{\partial x^i}\{a^i({\bf x})p({\bf x},t)\}.
\end{eqnarray}
This equation is deterministic because it does not involve any random fluctuations. Moreover, this equation is 
essentially equivalent to the ordinary differential equation, and therefore, it is origin of many equations used frequently 
in biology. For example, the control theory used for analyzing chemical-taxi of 
E.coli in \cite{Tau} is included in Eq. (\ref{eqn: determinisitic process}).

\paragraph{Diffusion process.} 
In the absence of the jump term (i.e., $W({\bf x}|{\bf y},t)=0$), the Differential Chapman-Kolmogorov 
equation reads as

\begin{eqnarray}
\frac{\partial p({\bf x},t)}{\partial t}&=&-\sum_{i=1}^N\frac{\partial}{\partial x^i}\{a^i({\bf x})p({\bf x},t)\}+\frac{1}{2}\sum_{i,j=1}^N\frac{\partial^2}{\partial x^i x^j}\{b^{ij}({\bf x})p({\bf x},t)\},
\end{eqnarray}
which is known as a diffusion process, which will be used later. 

\paragraph{Jump process.}
In contrast, if we assume that $a^i({\bf x})=b^{ij}({\bf x})=0$, 
the Differential Chapman-Kolmogorov 
equation takes the form

\begin{eqnarray}
\frac{\partial p({\bf x},t)}{\partial t}=\int dy[W({\bf x}|{\bf y},t)p({\bf y},t)-W({\bf y}|{\bf x},t)p({\bf x},t)],
\end{eqnarray}
which is known as a jump process. It means that the path trajectory of ${\bf X}_t$ will exhibit discontinuities 
(large jumps) at specific discrete
points.
 
It is also known that the jump processes can represent some kind of chemical reactions (Ref. \cite{Kampen}). Therefore, we may 
use this equation to analyze the metabolic pathways \cite{kegg} in cells, which are composed of chemical  
reactions and chemical compounds.

\paragraph{How to use the Differential Chapman-Kolmogorov equation.}

We can obtain the dynamics of probability density $p({\bf x}, t ~|~ {\bf x}_0, t_0)$ for any time $t$, from 
experimental data of instantaneous transition probability $T_\epsilon({\bf y},{\bf x})$ ($\epsilon$ is sufficiently small and fixed), by 
the following procedure:

\begin{itemize}

\item[{\it (i)}] Given the experimental data of instantaneous transition probability $T_\epsilon({\bf y},{\bf x})$ ($\epsilon$ is sufficiently small), we obtain $a^i({\bf x})$, $b^{ij}({\bf x})$ and $W({\bf x}|{\bf y},t)$ by using Eq. (\ref{eqn: initial condition1}), (\ref{eqn: initial condition2}) and (\ref{eqn: initial condition3}) respectively.\footnote{Notice that we do not need the whole data $T_\epsilon({\bf y},{\bf x})$ for any $\epsilon$. The necessary data is $T_\epsilon({\bf y},{\bf x})$ at a sufficiently small fixed $\epsilon$.} 

\item[{\it (ii)}] By inserting $a^i({\bf x})$, $b^{ij}({\bf x})$ and $W({\bf x}|{\bf y},t)$ into the Differential 
Chapman-Kolmogorov equation Eq. (\ref{eqn: master equation})  and by solving this 
PDE (Partial Differential Equation), we can obtain useful information like the distribution, the expectation value, the 
variance and the correlation at any time. 
\end{itemize}



\subsubsection{Initial instantaneous transition data $T_\epsilon({\bf y},{\bf x})$}

The initial instantaneous transition data $T_\epsilon({\bf y},{\bf x})$ of
{\it N} genes for studying correlations
should be determined by the experiment, which measures the short-time
transition
probability between the $N$ dimensional gene expression levels ${\bf x}$
at time $t$
and the $N$ dimensional gene expression levels ${\bf y}$ at time
$t+\epsilon$. Although some experiments have been done for measuring
gene expression time series of many organisms \cite{ueda,human,yeast}, the
statistics are not enough to completely determine the $T_\epsilon({\bf
y},{\bf x})$ and we believe that several experiments should be done under
the same
condition to have enough statistics. Moreover, the stochastic nature of the
fluctuations of the gene expression level, strongly supports
the idea of {\it many experiments-many genes} under the same conditions.

Therefore, for the time being, we assume that the expression of the initial
instantaneous
transition data $T_\epsilon({\bf y},{\bf x})$ of {\it N} genes correlation
is the Gaussian type,
which seems general enough to illustrate our model. The expression is as
follows;
\begin{eqnarray}\label{eqn:initial data T}
&&T_\epsilon({\bf y},{\bf x})\nonumber\\ 
&=&
\frac{1}{\sqrt{(2\pi\epsilon)^{N} det(\sigma^{ij})}} \\
&&\exp\Bigl[-\frac{1}{2}\epsilon^{-1}\sigma_{ij}(y^{i}-x^{i}-\epsilon\mu^i(m
^i-x^{i}))(y^{j}-x^{j}-\epsilon\mu^j(m^j-x^{j}) \Bigr],\nonumber
\end{eqnarray}
where $\sigma_{ij}$ is the inverse matrix of $\sigma^{ij}$ (i.e.
$\sum_k\sigma^{ik}\sigma_{kj}=\delta^i_j$). We note that the contraction of the indexes $i$, $j$ is done by using the Einstein summation convention.

This ITP $T_\epsilon({\bf y},{\bf x})$ (Eq. (\ref{eqn:initial data T})) allows us to analyze the gene correlation phenomena and, furthermore, 
has the mean reverting property, which is assumed in order to be
compatible with the observation in \cite{human,yeast} (see Fig. 1). Here, we
give some annotations on the parameters of $T_\epsilon({\bf y},{\bf x})$. $m^i$
denotes the average expression level of genes $i$ and $\mu^i$ means the
tendency to reverting to $m^i$. Finally, $\sigma_{ij}$ indicates the correlation
between gene $i$ and $j$. These parameters will be clear to the readers in the following
sections.

\subsubsection{Computation of $a^i({\bf x})$, $b^{ij}({\bf x})$ and $W({\bf x}|{\bf y},t)$}
In this section, we compute $a^i({\bf x})$, $b^{ij}({\bf x})$ and $W({\bf x}|{\bf y},t)$ from the initial 
data of ITP $T_\epsilon({\bf y},{\bf x})$. Inserting Eq. (\ref{eqn:initial data T}) into (\ref{eqn: initial condition1}), (\ref{eqn: initial condition2}) and (\ref{eqn: initial condition3}), we obtain
\begin{eqnarray}\label{eqn: initial condition}
a^i({\bf x})=\lim_{\epsilon \to 0}\frac{1}{\epsilon}\int_{|{\bf y}-{\bf x}|<\delta} (y^i-x^i) T_\epsilon({\bf y},{\bf x}) d{\bf y}=\mu^i(m^i-x^{i}),
\end{eqnarray}
and
\begin{eqnarray}\label{eqn: initial condition}
b^{ij}({\bf x})=\lim_{\epsilon \to 0}\frac{1}{\epsilon}\int_{|{\bf y}-{\bf x}|<\delta} (y^i-x^i)(y^j-x^j) T_\epsilon({\bf y},{\bf x}) d{\bf y}=\sigma^{ij},
\end{eqnarray}
and
\begin{eqnarray}\label{eqn: initial condition}
W({\bf y}|{\bf x},t)=\lim_{\epsilon \to 0} T_\epsilon({\bf y},{\bf x}) / \epsilon=0.
\end{eqnarray}

Here, we remark the following. The drift 
term $a^i({\bf x})=\mu^i(m^i-x^{i})$ represents 
that our model has the mean value property (i.e. the gene expression level of gene $i$ 
tends to revert the mean value $m^i$ with the quickness $\mu^i$. see Fig.
 1). The diffusion 
matrix $b^{ij}({\bf x})=\sigma^{ij}$ denotes the correlation between gene $i$ and $j$. Here, we remark 
that the correlation is constant in our model (Eq. (11)) for simplicity, although we 
could include the ${\bf x}$ dependence in the model in future work. Finally, the jump 
term $W({\bf x}|{\bf y},t)=0$ means that our model does not contain the jump process.



\subsubsection{Emergence of Kolmogorov equation and SPDE}
In the last section, we find out 
that $a^i({\bf x})=\mu^i(m^i-x^{i})$, $b^{ij}({\bf x})=\sigma^{ij}$ and $W({\bf x}|{\bf y},t)=0$. However, 
in order to keep the argument more general, we still 
consider the drift term $a^i({\bf x})$ and diffusion 
term $b^{ij}({\bf x})$ arbitrary while we assume that 
the jump term vanishes $W({\bf x}|{\bf y},t)=0$.

\paragraph{Kolmogorov Equation.}
If $W({\bf x}|{\bf y},t)=0$ vanishes, then the Differential 
Chapman-Kolmogorov equation (\ref{eqn: master equation}) becomes the Kolmogorov equation:       

\begin{eqnarray}
\frac{\partial p({\bf x},t)}{\partial t}=-\sum_{i=1}^N\frac{\partial}{\partial x^i}\{a^i({\bf x})p({\bf x},t)\}+\frac{1}{2}\sum_{i,j=1}^N\frac{\partial^2}{\partial x^i x^j}\{b^{ij}({\bf x})p({\bf x},t)\},
\end{eqnarray}
where $p({\bf x},t)=p({\bf x},t|{\bf x}_0,t_0)$.

\paragraph{SPDE.}
In addition, it is known that the Kolmogorov equation 
is equivalent to the following stochastic partial differential equation (SPDE):

\begin{eqnarray}\label{eqn: SPDE}
dX_t^{i}=\alpha^i({\bf X}_t) dt + \sum_{j=1}^N \beta^{ij}({\bf X}_t)dW_{j}(t).
\end{eqnarray}
Here, the multi-dimensional stochastic variable ${\bf X}_t$ denotes the gene expression level, $\alpha^i({\bf x})=a^i({\bf x})$ 
denotes the average change of the instantaneous 
transition of the gene expression level per unit time, $\beta^{ij}({\bf x})$ 
denotes the covariance of instantaneous transition of the gene expression level per unit time 
given by $b^{ij}({\bf x})=\sum_{k=1}^N\beta^{ik}({\bf
x})\beta^{jk}({\bf x})$ (Here we remark that $\beta^{ij}(x)$ is not 
uniquely determined from $b_{ij}(x)$ due to the rotation group ambiguity.) and ${\bf W}(t)=(W_{1}(t),\cdots,W_{N}(t))$ denotes 
the multi-dimensional Wiener process where all the processes are independent of each other $dW_{i}(t)dW_{j}(t)=\delta_{ij}dt$. 
We also note that the stochastic calculus in our approach follows the Ito rule by construction, and not the Stratonovich rule \footnote{See \cite{Kampen,Gard} for further discussions on the Ito-Stratonovich dilemma.}.

\paragraph{Correlated-SPDE.}

Although the above SPDE (\ref{eqn: SPDE}) is good for a theoretical study, it is difficult to analyze Eq. (\ref{eqn: SPDE}) directly, 
since Eq. (\ref{eqn: SPDE}) includes the multi-components processes of $W_{1}(t), \cdots ,W_{N}(t)$ and the ambiguity of $\beta^{ij}({\bf x})$ relative to the rotation group. Therefore, 
we transform the above SPDE (\ref{eqn: SPDE}) into a more convenient expression (i.e, where the correlation is explicitly manifested) 
as follows:
 
\begin{eqnarray}\label{eqn: SPDE2}
dX_t^{i}=\alpha^i({\bf X}_t) dt + \beta^i({\bf X}_t)dz^i(t),
\end{eqnarray}
and the correlation is given by 
\begin{eqnarray}\label{eqn: correlation}
dz^{i}(t)dz^{j}(t)=\rho^{ij}(x)dt,
\end{eqnarray}
where $\beta^i(x)=\sqrt{b^{ii}(x)}$, $\rho^{ij}(x)=\frac{b^{ij}(x)}{\sqrt{b^{ii}(x)}\sqrt{b^{jj}(x)}}$ and $(-1\le \rho_{ij}(x) \le 1)$.

It is important to note that this expression is more easy
to understand and deal with 
from the practical point of view, since the SPDE (\ref{eqn: SPDE2})
has only one Brownian motion (fluctuations) component $dz^{i}(t)$, while many components of Brownian motion appear as $dW^{1}(t).\cdots,dW^{N}(t)$ in (\ref{eqn: SPDE}). Furthermore, all the correlation information among genes are gathered in $dz^{i}(t)dz^{j}(t)=\rho^{ij}({\bf x})dt$ in Eq. (\ref{eqn: correlation}), which by the way 
is a more easy expression to deal with.

\subsubsection{Analysis of our model}
>From general analysis of Kolmogorov equation, we return to our original situation where the drift $a^i({\bf x})=\mu^i(m^i-x^i)$, the diffusion $b^{ij}({\bf x})=\sigma^{ij}$ and the jumping term $W({\bf x}|{\bf y},t)=0$. Then, by inserting them into the SPDE Eq. (\ref{eqn: SPDE2}) and (\ref{eqn: correlation}), we obtain   
\begin{eqnarray}\label{eqn: BS}
dX_t^{i}=\mu^i (m^i- X_t^{i}) dt + \sigma^i dz_t^{(i)},
\end{eqnarray}
where $\sigma^i=\sqrt{\sigma^{ii}}$ and the correlation is given by
\begin{eqnarray}\label{eqn: correlation of BS}
dz^{i}(t)dz^{j}(t)=\rho^{ij}dt=\frac{\sigma^{ij}}{\sigma^{i}\sigma^{j}}dt.
\end{eqnarray}

This reduced model is also known as Vasicek model in financial engineering (see Ref. \cite{black}). 
This SPDE directly gives useful information about the properties of the model as follows:

\begin{itemize}

\item[{\it (i)}] {\it Mean reverting}. The model (Eq. (\ref{eqn: BS})) has mean reverting property. We can observe this phenomena in Fig. 1, which is 
obtained from data of gene expression time series experiments of human and yeast organism \cite{human, yeast}. 

\item[{\it (ii)}] {\it Multi-correlation}. Eq. (\ref{eqn: BS}) also exhibits an embedded 
correlation relationship  given by Eq. (\ref{eqn: correlation of BS}). Therefore, by using this relationship, we can analyze  
the gene correlation phenomena among genes using our model.  



\end{itemize}

\subsubsection{Gene expression dynamical solution}

By using Ito formula, we can solve the SPDE (Eq. (\ref{eqn: BS})) and derive the dynamical solution of gene expression 
\begin{eqnarray}
X_t^{i}=m^i+(x_0^{i}-m^i)e^{-\mu^i t}+\sigma^i \int_0^t e^{-\mu^i (t-s)} dz^i(s),
\end{eqnarray}
where $x_0^{(i)}=X_0^{(i)}$ is the initial value of gene expression level. 
>From this solution, we can obtain the most important quantities which contain the relevant information of the system: 
expectation value and variance and covariance of multi-dimensional gene expression level at any time as follows: 
\begin{eqnarray}
E[X_t^{i}]=m^i+(x_0^{i}-m^i)e^{-\mu^i t},
\end{eqnarray}

\begin{eqnarray}
V[X_t^{i}]=\frac{(\sigma^i)^2}{2\mu^i}(1-e^{-2\mu^i t}),
\end{eqnarray}
and
\begin{eqnarray}
Cor[X_t^{i},X_t^{j}]=\frac{\sigma^i \sigma^j \rho^{ij}}{\mu^i+\mu^j}(1-e^{-(\mu^i+\mu^j)t}).
\end{eqnarray}

Roughly speaking, it means that if we specify the initial state ($x_0^{(i)}=X_0^{(i)}$) (for example, a patient receives a chemical treatment
 which modifies the amount of mRNA in cells), then  
we may predict the effect of that treatment in cells by knowing the 
expectation value, variance and covariance of multi-dimensional gene expression level at any time.

\subsection{Computation of simulated data}
In order to obtain the sample path 
of multi-gene correlation dynamics from Eq. (\ref{eqn: BS}), we use the difference 
equation corresponding to Eq. (\ref{eqn: BS}) as follows. For sufficiently large number $n$. We equally 
divide the time interval $[0,t]$ by $t_i=i\Delta t$ ($i=0,\cdots,n$ and $\Delta t=t/n$ ), then from Eq. (\ref{eqn: BS}) we obtain:
\begin{eqnarray}\label{eq: difference of vasicek}
X_{i+1}^{j}=X_i^{j}+\mu^j (m^j- X_i^{j}) dt + \sigma^j \Delta z_t^{(j)},
\end{eqnarray}
where
\begin{eqnarray}
X_i^{j}=X_{t_i}^{j}~~~~~~~\mbox{and}~~~~~~~
\Delta z_t^{(j)}=z_{t_{i+1}}^{(j)}-z_{t_{i}}^{(j)}\sim N_n(0,\rho^{ij}),
\end{eqnarray}
where $N_n(0,\rho^{ij})$ denotes the $n$-dimensional normal distribution and  
the mean vector is zero and the variance matrix is given by $\rho^{ij}$.
 We repeatedly use Eq. (\ref{eq: difference of vasicek}) to obtain the sample path. 
  In Fig. 2, we show the five sample paths for total positive correlation $\rho^{12}=1$, slightly positive correlation $\rho^{12}=0.5$, no correlation
  $\rho^{12}=0$, slightly negative correlation $\rho^{12}=-0.5$ and total negative correlation $\rho^{12}=-1$, respectively.

  Although the correlation $\rho^{ij}$ between genes is easy to be characterized by the theoretical analysis of simulated data of gene expression (Eq. (\ref{eqn:initial data T})), this issue becomes more difficult in a practical problem. Precisely, the most important experimental problem related to the analysis of time series of gene expression data by using Microarrays/GeneChips technologies is to identify significant correlations between observables of genes. The criteria that we may use for considering that two genes are significantly correlated is as follows: (1) By using a given time-series set of gene expression experimental data, our theory can be used to calculate the correlation value $\rho^{ij}$ for each couple of genes. In the case that this value is in the vicinity of the value one ($\rho^{ij}\simeq 1$) for two genes, we may consider that both genes are significant enough correlated.

  In addition, the following criterias should also be taken into account. (2) The correlation $\rho^{ij}$ between genes in our approach is obtained by inserting the ITP $T_\epsilon({\bf y},{\bf x})$ into the model. This ITP $T_\epsilon({\bf y},{\bf x})$ is obtained by using experimental techniques as Microarry/GeneChips. Currently, these technologies have a non-zero instrumental noise that in some cases may exceed 30\%. Therefore, it is important that this source of noise is reduced as much as possible for each experiment in order to evaluate with more accuracy the correlation between genes. Finally, (3) our theory for predicting the correlation phenomena is based on a stochastic approach. As we explained through the text, and in more extension in the next section, many experiments for analysing gene expression time series are encouraged to have a enough statistics to achieve a precise interpretation of the dynamics of gene correlation. Therefore, by increasing the number of the experiments, the confidence of the correlation observable $\rho^{ij}$ predicted by our theory would be improved.

\section{Experimental proposal}
The most important factor in our model is the initial 
data of ITP $T_\epsilon({\bf y},{\bf x})$, which characterizes the gene correlation system. 
However, as far as we know, we do not have enough experimental data for completely determining the ITP $T_\epsilon({\bf y},{\bf x})$. Therefore, to determine this ITP $T_\epsilon({\bf y},{\bf x})$, 
we propose an experiment, which measures the short-time transition 
probability between the $N$ dimensional gene expression levels ${\bf x}$ at time $t$ and 
the $N$ dimensional gene expression levels ${\bf y}$ at 
time $t+\epsilon$. The novelty, is that such experiments should be carried out at least hundred times to have enough statistics, and
 in addition, should be done under the same
external condition.

Interestingly, the stochastic nature of the
fluctuations of the gene expression level \cite{elo}, strongly supports
the idea of {\it many experiments-many genes} under the same conditions.

If we can obtain the experimental 
data of $T_\epsilon({\bf y},{\bf x})$, our model can predict the future behavior 
of multi-gene correlation using our construction (for example, the expectation value 
and variance of the gene expression level at any time in the future) from the 
initial value of gene expression levels. In particular, it would contribute 
to uncover how to regulate specific genes by applying some external action (e.g., a patient
under medical treatment).

\section{Conclusions}

We have carried out a theoretical study on gene expression correlation, which is one of the crucial topics of genomics 
in the current post-sequence era. Our study indicates that it is possible to analyze 
the dynamics underlying the gene expression correlation phenomena by using only one assumption
 the Markov property. In other words, it means that {\it the multi-dimensional 
 correlation dynamics of gene expression obeys the Markov property}. 

Our theoretical approach of multi-gene expression dynamics indicates that we can 
specify an initial state (${\bf X}_{t_0}={\bf x}_0$) of gene expression in a cell
and be able to predict the most relevant observables of the distribution of genes as expectation value, variance and 
covariance of multi-dimensional gene expression level at any time in the future. This feature represents 
an important step forward in the current
analysis of gene correlation analysis and have potential implications for genetic engineering, for example by developing personalized 
medicines according to the features of individuals.
 
Furthermore, in order to achieve the above described goals we presented an experimental proposal. The main idea 
of this new proposal is that
{\it many} experiments of {\it many} genes would be useful for completely uncover 
the dynamics of multi-genes in cells. 

It is also worth noticing that stochastic theory offers a huge and rich variety of tools for 
studying the gene expression fluctuations, and by
extension many other cellular phenomena. For example, it is known that the jump processes described by Eq. (8) 
can represent some kind of chemical reactions. Therefore, as future work we may use that equation 
to analyze the metabolic pathways in cells,
which are composed of chemical reactions and chemical compounds. 

The availability of complete
genomes for several organisms has definitely opened new and exciting possibilities of studying the gene correlation dynamics and mechanisms.
Consequently, we believe that our theoretical model, together with the experimental proposal, may further serve 
to understand the regulatory interactions among
genes and contribute to enlighten the advances of the post-sequence era.

\newpage



\begin{figure}[htb]
\setlength{\unitlength}{1cm}
\begin{picture}(15,12)(-1,-1)
\put(0,-1){\includegraphics[scale=0.65]{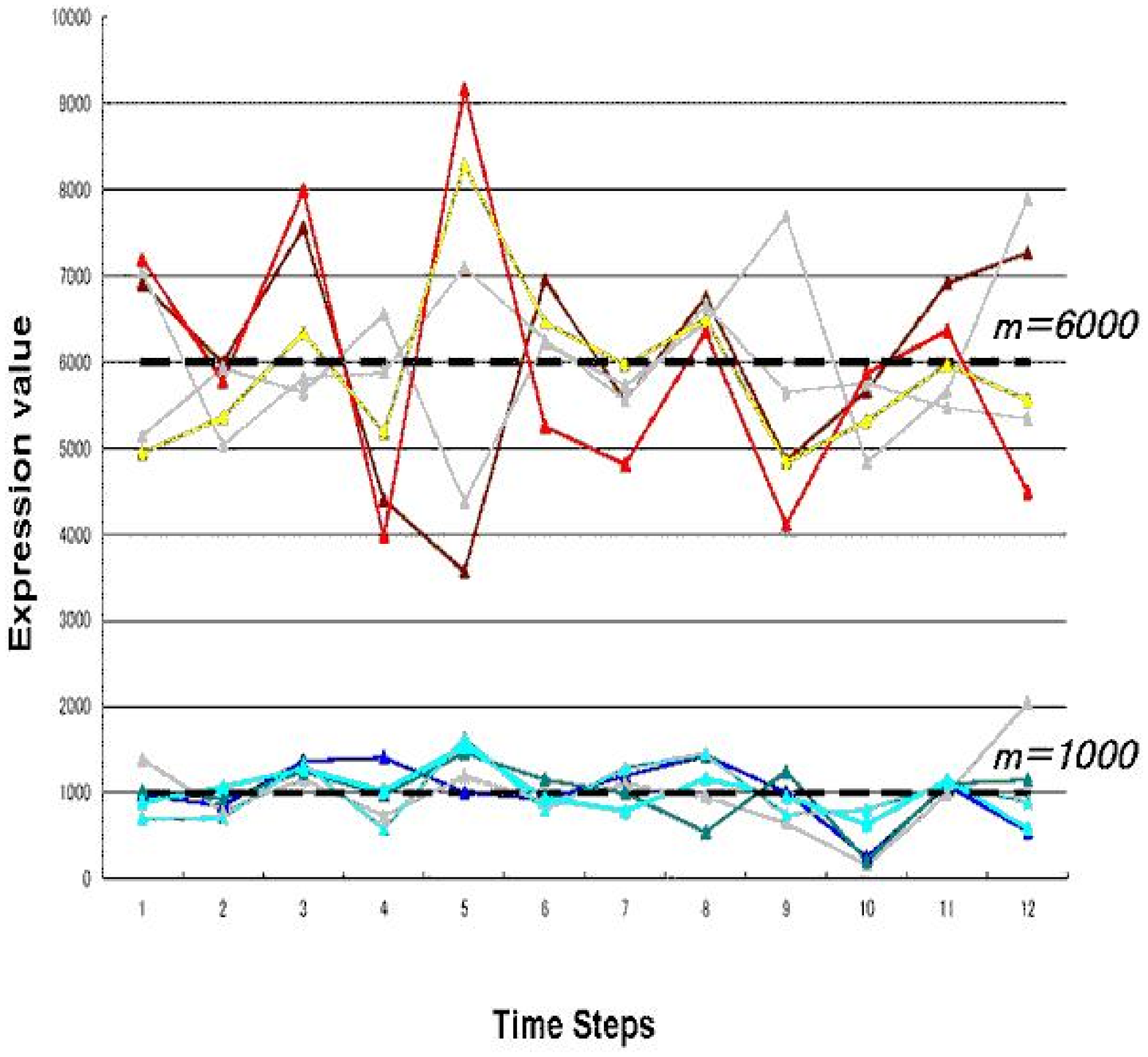}}
\end{picture} 
\caption{\small{We show the experimental absolute value of gene expression level (vertical axis) vs. time (horizontal axis) 
of a selected group of genes which belong to human organism \cite{human}. We see that the gene expression
value fluctuates around the mean value $m$=6000 and $m$=1000.}}
\label{fig: construction}
\end{figure}  




\begin{figure}[htb]
\setlength{\unitlength}{1cm}
\begin{picture}(15,12)(-1,-1)
\put(-2,0){\includegraphics[scale=0.7]{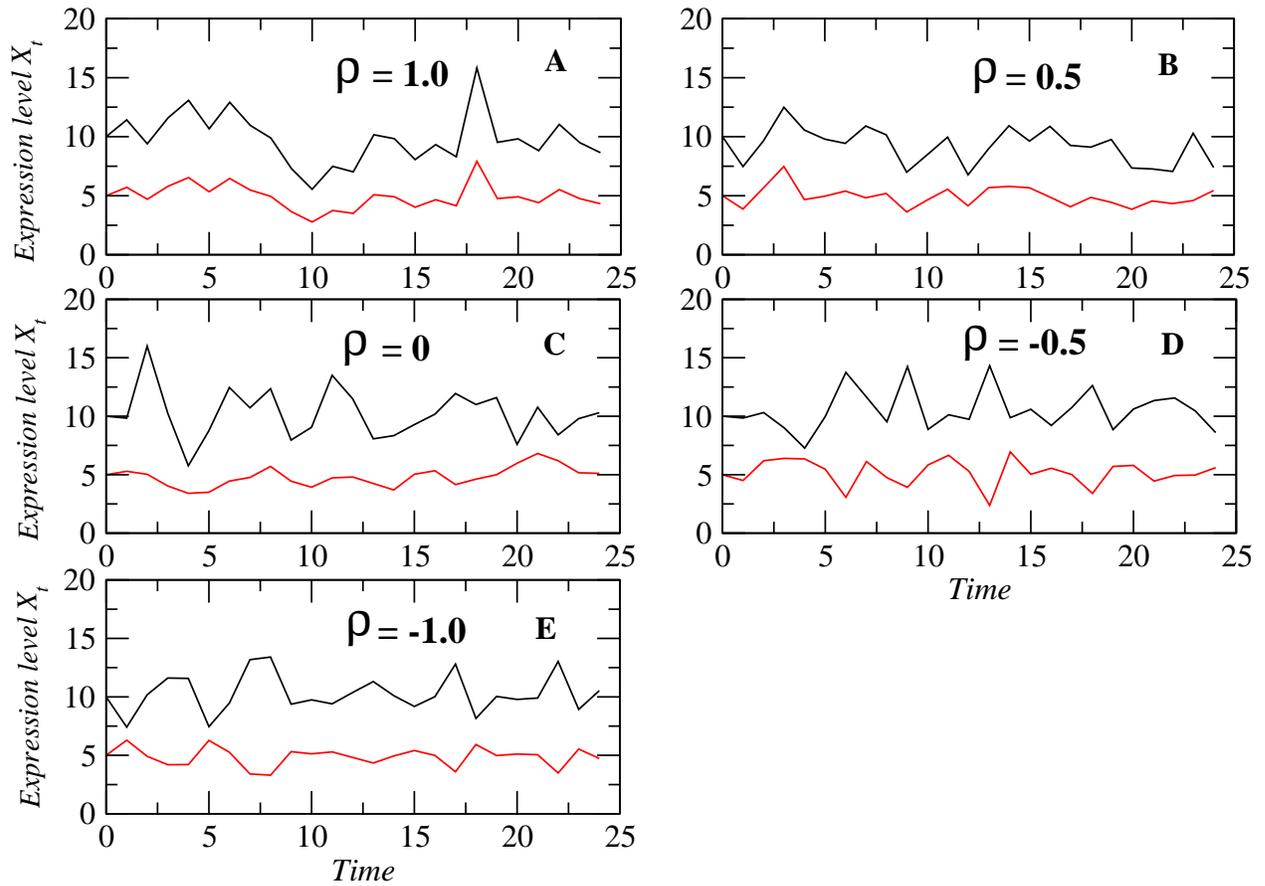}}
\end{picture}
\caption{\small{We show the simulated results of our model for different values of $\rho$. Values of $\rho$ are indicated 
in each figure, from $A$ to $E$. (A) $\rho=1.0$ indicates that both genes are totally positively correlated. (B) $\rho=0.5$ indicates that both genes are slightly positively correlated. (C) $\rho$=0 indicates 
uncorrelation between genes. (D) $\rho=-0.5$ indicates that both genes are slightly negatively correlated. (E) $\rho=-1.0$
indicates that both genes are completely negatively correlated.}}
\label{fig:Initial daata}
\end{figure}





\end{document}